\documentclass[amssymb,aps,showkeys,]{revtex4-2}

\usepackage{graphicx}
\usepackage{url}
\usepackage{multirow}
\usepackage{color}
\usepackage{amsmath}
\usepackage{subcaption}
\usepackage{pgfplots}

\usepackage{tikz}
\usetikzlibrary{shapes.geometric,arrows}
\usepackage{qcircuit}
\usepackage{braket}

\usepackage{comment}

\includecomment{sectiontocompile}

\usepackage{pgfplots}
\pgfplotsset{compat=newest}
\usepgfplotslibrary{groupplots}
\usepgfplotslibrary{dateplot}

\newcommand{\ds}{\displaystyle}
\renewcommand{\thesection}{\arabic{section}}
\renewcommand{\thetable}{\arabic{table}}
\renewcommand{\thesubsection}{\thesection.\arabic{subsection}}

\begin{document}

\title{Several fitness functions and entanglement gates in quantum kernel generation}

\author{Haiyan Wang}
\email{haiyan.wang@asu.edu}
\affiliation{School of Mathematical and Natural Science,  Arizona State University, Phoenix, AZ 85069, USA}

%\date{}

\begin{abstract}
Quantum machine learning (QML) represents a promising frontier in the quantum technologies. In this pursuit of quantum advantage, the quantum kernel method for support vector machine has emerged as a powerful approach. Entanglement, a fundamental concept in quantum mechanics, assumes a central role in quantum computing. In this paper, we investigate the optimal number of entanglement gates in the quantum kernel feature maps by a multi-objective genetic algorithm. We distinct the fitness functions of genetic algorithm for non-local gates for entanglement and local gates to gain insights into the benefits of employing entanglement gates.  Our experiments reveal that the optimal configuration of quantum circuits for the quantum kernel method incorporates a proportional number of non-local gates for entanglement.  The result complements the prior literature on quantum kernel generation where non-local gates were largely suppressed.  Furthermore, we demonstrate that the separability indexes of data can be leveraged to estimate the number of non-local gates required for the quantum support vector machine's feature maps. This insight can be helpful in selecting appropriate parameters, such as the entanglement parameter, in various quantum programming packages like https://qiskit.org/ based on data analysis. Our findings offer valuable guidance for enhancing the efficiency and accuracy of quantum machine learning algorithms.

\end{abstract}

\keywords{support vector machine; quantum feature map; genetic algorithm; entanglement gate}

\maketitle

%------------------- Section 1--------------------------

\section{Introduction}\label{sec1}
Quantum machine learning (QML) stands out as one of the most promising and fascinating applications of quantum technologies. In this pursuit of quantum advantage, the quantum kernel method via support vector machine has emerged as a powerful approach, drawing attention from various research groups \cite{BYH22,GGC22,LAT21,MBS22,PQW20,RML14,SCH21,SYG20, BWP17,SSP14,Schuld2022, huang2021,Di2023,HCT19,MS21}.   In \cite{HCT19}, two effective approaches were introduced for the construction of quantum SVMs: the Quantum Kernel Estimator and the Quantum Variational Classifier. These methods harness classical data and map it into the quantum state space via a quantum feature map. A pivotal element in this procedure is the quantum circuit, which facilitates the transformation of the dataset from its initial low-dimensional real space to a higher-dimensional quantum state space, commonly referred to as the Hilbert space 

The choice of an appropriate feature map, matched to a suitable kernel, emerges as a decisive factor in the success of kernel methods. The selection of the right feature map becomes particularly critical depending on the dataset under consideration.  Genetic algorithms, rooted in the principles of natural selection, naturally emerge as a valuable tool for classical support vector machines \cite{Ji2020}, and have been applied to quantum computing in a considerable number of works \cite{Lahoz-Beltra2016,Acampora2021,Chen_2022,CYH16} and quantum support vector machines \cite{Kavitha2022,ARG21,CC2022}. \cite{ARG21} first proposed the genetic algorithm for the auto generation of feature map in quantum support vector machine.  Additional advancements \cite{baran2021, CSU20} have introduced the automatic generation of quantum circuits through the utilization of multi-objective genetic algorithms.  These algorithms provide a potent avenue for optimizing gate or circuit structures, circumventing common challenges like local minima and barren plateaus \cite{li2017, lamata2018, chivilikhin2020}.

Entanglement, a fundamental concept in quantum mechanics, plays a pivotal role in enabling quantum computing.  In their work, \cite{Nguyen2022} presented an automated search algorithm aimed at determining the optimal layout for entanglement in supervised quantum machine learning. They accomplished this using sequential model-based optimization (SMBO). Their empirical findings demonstrate that the quantum embedding architecture produced by the automated search algorithm surpasses manually designed architectures in terms of predictive performance.

In this paper, we investigate the optimal number of entanglement gates in the quantum feature maps for quantum support vector machine by multi-objective genetic algorithm. We use three fitness functions that simultaneously maximizes classification accuracy while minimizing the gate costs of the quantum feature map's circuit. We separate the fitness functions for local gates and non-local (controlled-NOT (CNOT) ) gates. By doing so, we obtain a set of non-dominated points that reveal a more balanced configuration of quantum circuits for the quantum kernel method, incorporating a proportional number of CNOT gates to facilitate entanglement.  This result complements 
the previous works \cite{ARG21,CC2022} where their design of fitness functions has predominantly led to a suppression of entanglement gates. In addition, the experiments in this paper scale well for larger qubits than the previous work in \cite{Nguyen2022} where their experiment requires a significantly larger amount of computational resources.

The results of this paper indicate that various separability indexes can be used to help assessing the construct efficient quantum circuits for achieving high accuracy. Currently, the documentations for quantum kernels at https://qiskit.org/ do not provide a guide for how these parameters should be chosen. For example, the entanglement parameter of ZZFeatureMap has three options: linear, full and circular. The results of this paper can be used to help choosing these parameters based on analysis of data.     

Key new contributions of this paper are 1) using three fitness functions for a genetic algorithm to show that the optimal quantum circuits for the quantum kernel map incorporates a proportional number of entanglement gates, which complements the result in \cite{ARG21,CC2022}; 2) experimentally demonstrating that the separability indexes of data could be used to estimate the number of entanglement gates based on analysis of data.

\section{Support Vector Machine with Quantum Kernels}\label{sec2}

\subsection{Support vector machine}\label{sec2-1}

Support vector machines (SVMs) are a supervised learning method used for classification and regression. The goal is to find the optimal hyperplane that separates two classes by maximizing the margin. Mathematically, for a given labeled training set ${(x_i, y_i)}$ where $x_i$ are the features and $y_i$ are the labels (+1 or -1), the optimization problem is:

\begin{align*}
& \text{minimize} \quad ||w||^2 \
& \text{subject to} \quad y_i(w \cdot x_i + b) \geq 1, \quad \forall i
\end{align*}
where $w$ is the normal vector to the hyperplane, $b$ is the bias term, and the constraints ensure we maximize the margin. This optimization finds the maximum margin hyperplane.  The $x_i$ points nearest the hyperplane are called support vectors.  Overall, SVMs are powerful for complex classification and regression tasks. The primal optimization problem can be alternatively represented as the Lagrange dual problem:

\begin{equation} \label{L_dual}
    \begin{array}{cl}
        \max &  L(\alpha) = \ds \sum_{i=1}^{n}\alpha_i - \frac{1}{2}\sum_{i=1}^{n}\sum_{j=1}^{n}
        \alpha_i\alpha_jy_iy_jx_i \cdot x_j \\
        \text{s.t.} & \ds \sum_{i=1}^{n} \alpha_i y_i=0,\ \alpha_i \geq 0
    \end{array}
\end{equation}\medskip

Given a function $\Phi$, which maps a data point  $x$  to higher dimensional space, the above problem can be transformed into  

\begin{equation} \label{L_dual}
    \begin{array}{cl}
        \max &  L(\alpha) = \ds \sum_{i=1}^{n}\alpha_i - \frac{1}{2}\sum_{i=1}^{n}\sum_{j=1}^{n}
        \alpha_i\alpha_jy_iy_j\Phi(x_i) \cdot \Phi(x_j) \\
        \text{s.t.} & \ds \sum_{i=1}^{n} \alpha_i y_i=0,\ \alpha_i \geq 0
    \end{array}
\end{equation}\medskip
The kernel function, denoted as $K(x_i, x_j) = \Phi(x_i)\cdot \Phi(x_j)$, enables SVM computations by only requiring knowledge of how to compute the inner product of $\Phi(x_i)$ and $\Phi(x_j)$, rather than explicit knowledge of the feature map $\Phi$.  Kernel methods encompass a collection of pattern analysis algorithms that rely on the utilization of kernel functions, functioning within high-dimensional feature spaces.  To handle non-linear problems, SVMs use kernel functions to map the data to a higher dimensional space where separation is possible.  The kernel function operates by implicitly mapping input data into these higher-dimensional spaces, thereby simplifying the problem-solving process. In essence, kernels possess the ability to transform originally non-linearly separable data distributions into linearly separable ones, a phenomenon commonly referred to as the "kernel trick."

\subsection{Quantum kernel method}\label{sec2-2}

In the context of quantum computing, the Quantum Kernel Estimator detailed in \cite{HCT19} presents effective strategies for constructing a quantum SVM.  A significant challenge in quantum machine learning lies in the effective encoding of classical information into quantum states for quantum computing.  \cite{HCT19} achieves this by encoding data within the quantum state space through a quantum feature map. The choice of the feature map is a pivotal decision, often contingent on the characteristics of the dataset to be classified. It is worth noting that the feature map's significance extends beyond utilizing quantum state space as a feature space; it also involves the manner in which data is mapped into this high-dimensional space.

Through the utilization of quantum circuits, a data point $\mathbf{x}=(x_j) \in \mathbb{R}^n$ is transformed into an n-qubit quantum feature state, $|\Phi(\mathbf{x}\rangle\langle\Phi(\mathbf{x})|$.  This naturally leads to the definition of the kernel as $K(\mathbf{x},\mathbf{z})= | \langle \Phi (\mathbf{x}) \vert \Phi (\mathbf{z}) \rangle |^2 $.  The kernel method can be seamlessly integrated into quantum computing by through a unitary operator $\mathcal{U}_{\Phi(\mathbf{x})}$ applied to the initial state $\vert 0 \rangle ^n$,  $\vert \Phi (\mathbf{x}) \rangle = \mathcal{U}_{\Phi(\mathbf{x})} \vert 0 \rangle ^n$.  This approach facilitates the mapping of data points into the quantum Hilbert space ~\cite{SK19}, thereby leveraging the substantial advantages of quantum computing.

\begin{equation}\label{K_circuit}
    K(\mathbf{x},\mathbf{z}) = |\langle\Phi(\mathbf{x})|\Phi(z)\rangle|^2
    = |\langle 0^n|\mathcal{U}_{\Phi(\mathbf{x} }^\dagger)\mathcal{U}_{\Phi(\mathbf{z})}|0^n\rangle|^2
\end{equation}\medskip

%$\Phi(\mathbf{x}) = \mathcal{U}(\mathbf{x})|0^n\rangle\langle 0^n|\mathcal{U}^\dagger(\mathbf{x})$, achieved via a unitary circuit $\mathcal{U}(\mathbf{x})$. This circuit serves as a powerful tool for evaluating the kernel as follows:
%
%
%The quantum kernel estimation process entails evolving the initial state $|0^n\rangle$ on a quantum computer using $\mathcal{U}^\dagger(x_i)\mathcal{U}(x_j)$ and subsequently measuring the frequency of the outcome $0^n$. As long as the computation of the quantum kernel $K(x_i,x_j)$ remains efficient, we can conduct optimization tasks within the Hilbert space, with other optimization steps executed on a classical computer. For a more in-depth understanding of quantum kernel estimation, please consult the detailed information provided in~\cite{HCT19}.
%
%
%
%
%Developing feature maps based on quantum circuits that defy classical simulation is a crucial stride towards achieving a quantum advantage over classical methodologies. In this context, the work presented in \cite{HCT19} introduces a family of feature maps, conjectured to be classically hard to simulate, while being amenable to implementation as short-depth circuits on near-term quantum hardware. However, certain applications may warrant consideration of a more generalized form for the feature map.

In \cite{HCT19}, various widely-recognized feature maps like ZFeatureMap and ZZFeatureMap are proposed. One approach to customization involves employing a Pauli feature map and specifying a set of Pauli gates, deviating from the default $Z$ gates. For example, a representative form of our tailored quantum feature map might take the following structure:

$$
\mathcal{U}_{\Phi(\mathbf{x})}=\left(\exp \left(i \sum_{j, k} \phi_{\{j, k\}}(\mathbf{x}) Z_{j} \otimes Z_{k}\right) \exp \left(i \sum_{j} \phi_{\{j\}}(\mathbf{x}) P_{j}\right) H^{\otimes n}\right)^{d}
$$ 
where $P_j$ can be one of the rotation gates $I, X, Y, Z$.  The depth $d=1$ version of this quantum circuit is shown in Table \ref{he_moon23} below for $n=2$ qubits, and 
the coefficient $\phi_{S}(x)$, where $S$ is chosen to represent a set consisting of independent qubits or qubit pairs, e.g.,
\begin{equation}
\phi_S: x \mapsto
\begin{cases}
x_{i} & \text{if } S = \{i\} \\
(\pi - x_{i})(\pi - x_{j}) & \text{if } S = \{i, j\}
\end{cases}
\label{eq56}
\end{equation}
to encode the data.  

\begin{table}[h]
\[
\begin{array}{c}
\Qcircuit @C=1em @R=.7em {
\lstick{\ket{0}} & \gate{H} & \gate{R} & \ctrl{1} &  \qw  & \ctrl{1}&\qw \\
\lstick{\ket{0}} & \gate{H} & \gate{R} & \targ & \gate{R} & \targ & \qw
}
\end{array}
\]
\caption{Quantum circuit for $n=2$ qubits }  
\label{he_moon23} 
\end{table}
\noindent The circuit in Table \ref{he_moon23} involves a layer of Hadamard gates $H^{\otimes 2}$, followed by a subsequent layer of single-qubit gates $R=e^{i \phi_{{j}}(\mathbf{x}) P_{j}}$. These $R$ gates are characterized by a set of angles $\phi_{{j}}(\mathbf{x})$, each associated with a distinct axis, and $\phi_{{j}}(\mathbf{x})$ is contingent on the feature data. The diagonal entangling gate $e^{i \phi_{\{0,1\}}(\mathbf{x}) Z_{0} \otimes Z_{1}}$ is parametrized by an angle $\phi_{\{0,1\}}(\mathbf{x})$ and can be implemented using two CNOT gates and one $R=e^{i \phi_{\{0,1\}}(\mathbf{x}) Z_{1}}$ gate as shown in Table \ref{he_moon23}, here CNOT gate is used to establish entanglement among qubits. Specifically, the quantum gate is acted on 2 qubits, which has matrix representation:

$$
\mathrm{CNOT}=\left[\begin{array}{llll}
1 & 0 & 0 & 0 \\
0 & 1 & 0 & 0 \\
0 & 0 & 0 & 1 \\
0 & 0 & 1 & 0
\end{array}\right]
$$

The first qubit serves as the control qubit, while the second one is designated as the target qubit. If the control qubit is in the state $|0\rangle$, the target qubit retains its original state. Conversely, when the control qubit is in the state $|1\rangle$, the target qubit undergoes a bitwise flip. The classical counterpart of the CNOT gate corresponds to the reversible XOR gate.

Designing an appropriate kernel function for diverse datasets presents a formidable challenge. In our research, we employ a genetic algorithm to implement a global optimization approach for refining the structure of the quantum circuit. This methodology has proven to be effective in overcoming typical obstacles, such as barren plateaus, often encountered in conventional optimization techniques. As a result, it yields high-quality learning kernels \cite{ARG21, GA11}.

%Note that the ZZFeatureMap they provided
%is an unitary circuit defined as
%$U=\widetilde{U}_{\Phi(x)}H^{\otimes2}\widetilde{U}_{\Phi(x)}H^{\otimes2}$, where
%$$\widetilde{U}_{\Phi(x)}=\exp (i\Phi_1(x)ZI+i\Phi_2(x)IZ+i\Phi_{1,2}(x)ZZ)$$, is
%only suitable for a specific dataset. 

%-------------------- Section 3---------------------

\section{Evolutionary Multi-objective Optimization} \label{sec3}
Genetic algorithms belong to the category of optimization methodologies that draw inspiration from the mechanisms of evolution. These algorithms traverse a solution space by iteratively refining a population of individuals. Over successive generations, genetic operations dictate the choice of offspring aimed at enhancing one or more objectives. This evolutionary dynamic effectively culminates in the selection of highly suitable individuals from within a vast configuration space.

In the process of generating new solutions, a pair of "parent" solutions is chosen from a previously selected pool. Through techniques like crossover and mutation, a "child" solution is then created, inheriting many characteristics from its "parents." New pairs of parents are selected for each child, and this cycle continues until a fresh population of solutions, of the desired size, is formed. The effectiveness and applicability of a genetic algorithm depend significantly on the choice of genetic operations. Among these, the selection operator assumes a pivotal role as it determines a subset of the existing population to form a new generation through crossover and mutation operations. Mutation introduces random alterations to selected individuals' information, facilitating exploration of distant regions within the solution space. In contrast, crossover enables more radical exploration by exchanging genetic information between two individuals. Notably, mutation and crossover probabilities are fixed values, fine-tuned for optimal performance, while the selection probability is directly proportional to individuals' fitness.

To ensure the genetic algorithm's efficiency and convergence, early stopping conditions are employed to assess whether the evolutionary process has achieved its objectives. These conditions may involve monitoring the convergence or saturation of the fitness objective, maintaining a minimum accuracy threshold to continue the process, or setting a maximum number of generations. By leveraging these essential components, genetic algorithms emerge as valuable tools for addressing intricate optimization tasks across various domains.

An Evolutionary Multi-Objective Optimization (EMO) problem entails multiple objective functions that are subject to either minimization or maximization. Similar to single-objective optimization problems, multi-objective optimization problems may also incorporate various constraints that must be adhered to by all feasible solutions, including the optimal ones. For further insights, refer to \cite{ADB14, Miettinen1999, Deb2002}.

%Given the potential for both objective minimization and maximization, we present the multi-objective optimization problem in its general form:
%$$
%\left.\begin{array}{rll}
%\text { Minimize/Maximize } & f_{m}(\mathbf{x}), & m=1,2, \ldots, M ; \\
%\text { subject to } & g_{j}(\mathbf{x}) \geq 0, & j=1,2, \ldots, J ; \\
%& h_{k}(\mathbf{x})=0, & k=1,2, \ldots, K; \\
%& x_{i}^{(L)} \leq x_{i} \leq x_{i}^{(U)}, & i=1,2, \ldots, n .
%\end{array}\right\}
%$$

%A solution $\mathbf{x} \in \mathbf{R}^{n}$ is a vector of $n$ decision variables: $\mathbf{x}=\left(x_{1}, x_{2}, \ldots, x_{n}\right)^{T}$. The solutions satisfying the constraints and variable bounds constitute a feasible decision variable space $S \subset \mathbf{R}^{n}$. One of the striking differences between single-objective and multi-objective optimization is that in multi-objective optimization the objective functions constitute a multi-dimensional space,  in addition to the usual decision variable space. This additional $M$-dimensional space is called the objective space, $Z \subset \mathbf{R}^{M}$. For each solution $\mathbf{x}$ in the decision variable space, there exists a point $\left.\mathbf{z} \in \mathbf{R}^{M}\right)$ in the objective space, denoted by $\mathbf{f}(\mathbf{x})=\mathbf{z}=\left(z_{1}, z_{2}, \ldots, z_{M}\right)^{T}$. To make the descriptions clear, we refer a 'solution' as a variable vector and a 'point' as the corresponding objective vector.

\tikzstyle{startstop} = [rectangle, rounded corners, 
minimum width=3cm, 
minimum height=1cm,
text centered, 
draw=black, 
fill=red!30]

\tikzstyle{process} = [rectangle, 
minimum width=3cm, 
minimum height=1cm, 
text centered, 
text width=3cm, 
draw=black, 
fill=orange!30]

\tikzstyle{processY} = [rectangle, 
minimum width=3cm, 
minimum height=1cm, 
text centered, 
text width=3cm, 
draw=black, 
fill=yellow!30]

\tikzstyle{description} = [rectangle, 
minimum width=2cm, 
minimum height=1cm, 
text centered, 
text width=2cm, 
draw=black, 
fill=orange!30]

\tikzstyle{decision} = [diamond, 
minimum width=2cm, 
minimum height=1cm, 
text centered, 
draw=black, 
fill=green!30]
\tikzstyle{arrow} = [thick,->,>=stealth]

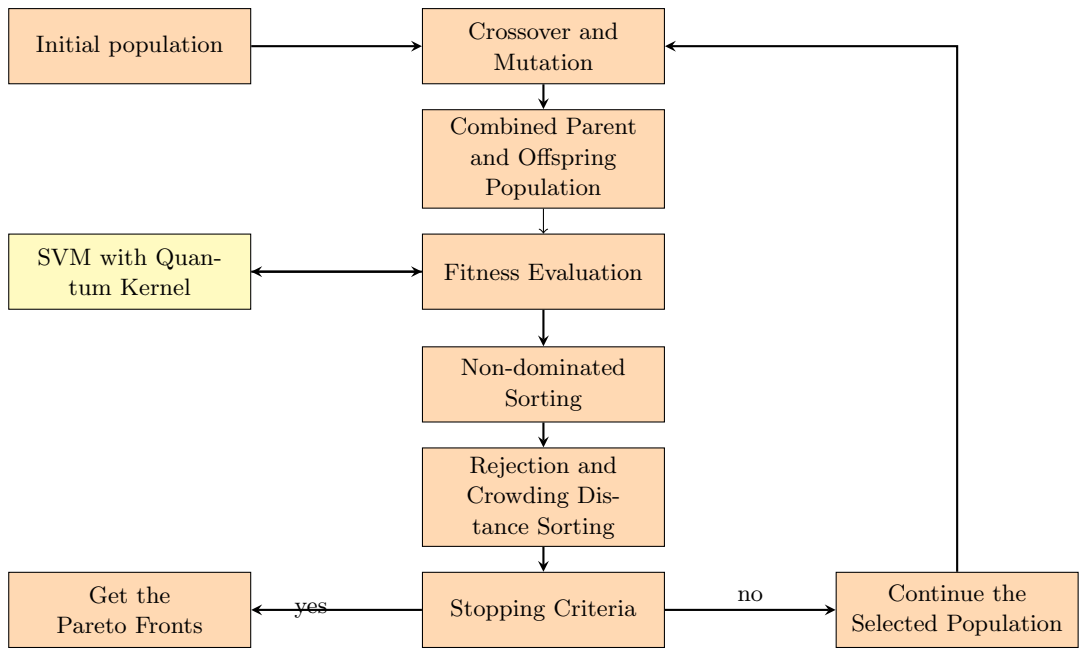
\begin{figure}
\centering

\begin{tikzpicture}[node distance=0.5cm]

\node (1) [process] {Initial population};
\node (2) [process, right of=1, xshift=5cm] {Crossover and Mutation};
\node (4) [process, below of=2, yshift=-1cm] {Combined Parent and Offspring Population};
\node (5) [process, below of=4,yshift=-1cm] {Fitness Evaluation};
\node (6) [process, below of=5, yshift=-1cm] {Non-dominated Sorting};
\node (7) [process, below of=6, yshift=-1cm] {Rejection and Crowding Distance Sorting};
\node (8) [process, below of=7, yshift=-1cm] {Stopping Criteria};
\node (9) [process, left of=8, xshift= -5cm] {Get the Pareto Fronts};
\node (10) [process, right of=8, xshift=5cm] {Continue the Selected Population};
\node (11) [processY, left of=5, xshift=-5cm] {SVM with Quantum Kernel};

%\node (pro1) [process, below of=in1] {Selection of Individual};
%\node (pro2b) [process, right of=dec1, xshift=2cm] {Termination Condition};
%\node (out1) [io, below of=pro2a] {Output};
%\node (stop) [startstop, below of=out1] {Stop};

\draw [arrow] (1)--(2) ;
\draw [arrow] (2)--(4) ;
\draw [->](4) -- (5);
\draw [arrow] (5) -- (6);
\draw [arrow] (6) -- (7);
\draw [arrow] (7) -- (8);
\draw [arrow] (8) -- node[anchor=east] {yes} (9);
\draw [arrow] (8) -- node[anchor=south] {no} (10);
\draw [arrow] (10) |- (2);

\draw [arrow] (5) -- (11);
\draw [arrow] (11) -- (5);

%\draw [arrow] (dec1) -- node[anchor=east] {yes} (pro2a);
%\draw [arrow] (dec1) -- node[anchor=south] {no} (pro2b);
%\draw [arrow] (pro2b) |- (pro1);
%\draw [-] (pro2a) -- (out1);
%\draw [arrow] (out1) -- (stop);
\end{tikzpicture}
  \caption{Non-dominated Sorting Genetic Algorithm (NSGA-II)}
  \label{NSGA-II}
\end{figure}

For multi-objective optimization, optimal solutions are characterized by a mathematical concept known as partial ordering, and this concept is described using the term "domination." In this paper, our focus is primarily on addressing unconstrained optimization problems, which are free from any equality, inequality, or bound constraints.  The notion of domination between two solutions is articulated as follows: Solution $\mathbf{x}^{(1)}$ is deemed to dominate another solution, $\mathbf{x}^{(2)}$, only if both of the following conditions hold true:

\begin{enumerate}
  \item The solution $\mathbf{x}^{(1)}$ is no worse than $\mathbf{x}^{(2)}$ in all objectives (for minimization problem, each component  of $\mathbf{x}^{(1)}$ is less than or equal to that of $\mathbf{x}^{(2)}$ ).  Consequently, the comparison between these solutions is made by evaluating their objective function values or by examining the positions of the corresponding points $\left(\mathbf{z}^{(1)}\right.$ and $\mathbf{z}^{(2)}$ ) within the objective space.
  \item The solution $\mathbf{x}^{(1)}$ is strictly better than $\mathbf{x}^{(2)}$ in at least one objective (for minimization problem, one component of $\mathbf{x}^{(1)}$ is strictly less than the corresponding component of $\mathbf{x}^{(2)}$, in addition to item 1)
\end{enumerate}

Points that remain not dominated by any other members within a set are termed as non-dominated points of class one, or simply non-dominated points. One distinctive feature of any two such points is that improvements in one objective only occur at the expense of at least one other objective—a trade-off relationship. This trade-off property among non-dominated points is interesting as it encourages the exploration of a diverse array of such points before reaching a final decision.  If the set of points contains all possible points within the search space, the points residing on the non-domination front, by definition, are immune to domination by any other point within the objective space. Consequently, they represent the Pareto-optimal points, collectively constituting the Pareto-optimal front.

The Non-dominated Sorting Genetic Algorithm (NSGA-II) was initially designed, in response to the limitations of early evolutionary algorithms such as the lack of elitism,  as outlined by Deb in \cite{Deb2002}, stands as one of the widely adopted Evolutionary Multi-Objective Optimization (EMO) techniques. The procedure of NSGA-II is illustrated in Figure \ref{NSGA-II} and can be summarized as follows. 

\begin{itemize}
    \item \textbf{Initial Population}: The process starts with the creation of an initial population of potential solutions. This population represents a set of candidate solutions to the optimization problem.
    
    \item \textbf{Crossover and Mutation}: Genetic operations such as crossover (recombination of individuals) and mutation (introducing small random changes) are applied to the individuals in the population. These operations mimic the principles of natural selection and genetic evolution.
    
    \item \textbf{Combined Parent and Offspring Population}: The results of the crossover and mutation operations are combined with the parent population to form a new generation of potential solutions. This process is essential for exploring a broader solution space.
    
    \item \textbf{Fitness Evaluation}: The fitness of each individual in the combined population is evaluated. In multi-objective optimization, fitness may be defined with respect to multiple objectives, seeking solutions that optimize several criteria simultaneously.
    
    \item \textbf{Non-dominated Sorting}: The individuals are sorted based on their dominance. In this step, solutions are categorized into non-dominated fronts, where no solution is worse than another in all objectives. These fronts help identify the Pareto-optimal solutions, which represent the best trade-offs between objectives.
    
    \item \textbf{Rejection and Crowding Distance Sorting}: Solutions that do not belong to the Pareto-optimal fronts are rejected. Additionally, a crowding distance sorting mechanism is used to further differentiate solutions within the same front, favoring diverse and well-distributed solutions.
    
    \item \textbf{Stopping Criteria}: A check is made to determine if the specified number of generations has been reached. If not, the algorithm proceeds to the next generation. If the termination condition is met, the algorithm stops. To avoid over fitting, comparisons with the baseline methods are also implemented. 
    
    \item \textbf{Get Pareto Fronts}: At the end of the algorithm, the Pareto fronts represent the optimal solutions that achieve a balance between conflicting objectives. These solutions offer decision-makers a range of trade-off options.
    
    \item \textbf{Continue with Selected Population}: Depending on the problem and its requirements, the selected population of solutions can be used as a starting point for further analyses, optimizations, or decision-making processes.
    
    \item \textbf{SVM with Quantum Kernel}:  For our work,  we use a Support Vector Machine (SVM) with a quantum kernel feature map to evaluate the fitness of solutions. The quantum kernel map is used to capture complex data relationships.
\end{itemize}

The resulting Pareto fronts for quantum feature maps serve as clear visual representations of the methodology for achieving high accuracy while keeping gate costs minimal. These Pareto-optimal fronts provide a useful means to analyze cost-effective trade-off strategies, especially in cases where numerous objectives might conflict with one another.  While we only consider the gate costs associated with the maximum accuracy in this paper, further examination of the three-dimensional Pareto-optimal fronts and their projections onto two-dimensional spaces could offer deeper insights into the significant influence of entanglements on quantum kernel accuracy.

\section{Quantum Kernel for Support Vector Machine}\label{sec4}

\subsection{Genetic quantum feature map}\label{featureMap}

Based on the method to generate quantum feature maps with a multi-objective optimization process in \cite{ARG21}, here in this paper, we choose to implement the NSGA-II algorithm, which autonomously generates quantum feature maps for support vector machines.  As we discuss before, the NSGA-II algorithm uses elitism and other features to maintain diversity within the Pareto set.  It actively seeks out circuits that, once trained through a support vector machine, maximize accuracy, all while striving to minimize circuit complexity. This complexity is measured in terms of local gates (including Hadamard gates and rotation gates) and CNOT gates.  The algorithm's selection process for individuals advancing to the next generation depends on Pareto-dominance and density-based metrics as illustrated in Figure \ref{NSGA-II}. 

In this paper,  we separate CNOT gates from local gates and add additional optimizing functions. We modified the encoding method \cite{ARG21,CC2022} to create a quantum circuit with a minimum of local and CNOT gates and maximum of accuracy.  To simplify the problem, we choose the feature number as the number of qubits ($N$).  The encoding scheme in this paper begins with $N$ number of $H$ gates and $N$ possible rotation gates and extra two bits to indicate a rotation with respect to $X,Y,Z$, followed by possible $H$ entanglement which can be implemented by two rotation gates and a CNOT gates as shown in Table \ref{he_moon23}, the last two bits are reserved for repetitions. The encoding bit string is of size , $N (\text{ rotation for each qubit})+\binom{N}{2}(\text{ number of possible CNOT gates}) +2(\text{ type of rotation})+2(\text{ repetition})$. We also include the data into the encoding.  Because the feature number is the same as the number of qubits,  thus the $i$-th gene that operates on the qubits of the quantum register possibly depends on the $i$-th variable from the input data $\mathbf{x}\in \mathbb{R}^n$ as in Equation (\ref{eq56}).

%Specifically, rotation gate is  parameterized by a value in the input data.
%$R_{\sigma}(\mathbf{x})=\exp(i\phi(\mathbf{x})\sigma)$
%where $\sigma^\alpha$ are  one of Paulis  = $[X, Y, Z]$. As we see from Figure \ref{he_moon23},  an entanglement between circuits $i,j$ can be implemented using two CNOT gates and rotation $e^{i \phi_{\{0,1\}}(x) Z_{1}}$ gate. For the rotation gate of the entanglement implementation between circuits $i,j$. The rotation parameter is chosen to be 
%$R_Z(i, j)=\exp(i \phi_{i, j}(\mathbf{x}) Z).$ 

\begin{figure}[htbp]
    \centering
    \includegraphics[width=0.8\textwidth]{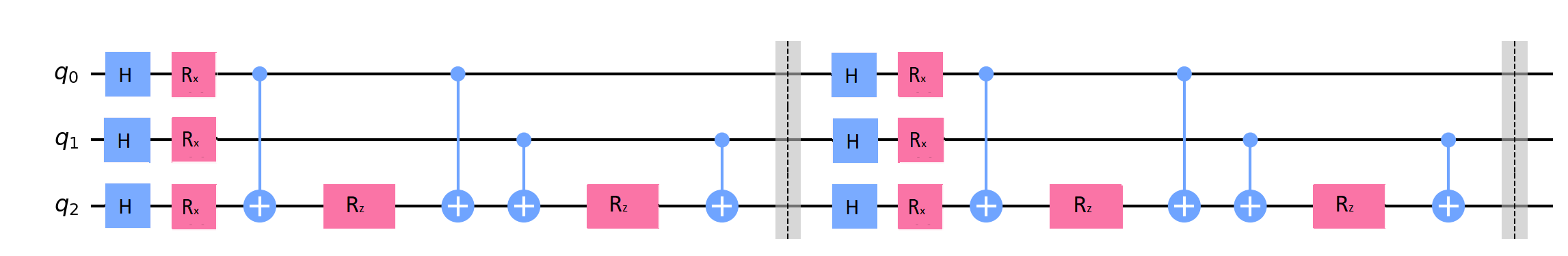}
    \caption{Example for the encoding scheme:  the number of qubits is 3,  the binary bits are [1, 1, 1, 1, 0, 0, 1, 1, 1, 0],  the first three bits for 3 rotation gates, and next two bits indicates the rotation with respect to $X, Y, Z$, followed 3 bits for possible entanglements,  the last two bits are for repetitions. }
    \label{encode}
\end{figure}

The genetic algorithm randomly initiates the initial population of individuals. The evaluation function deciphers each individual, constructing an associated quantum circuit. This circuit, in conjunction with the training dataset, is employed to assess the corresponding kernel matrices, which are then integrated into SVM algorithms for fitness function computation.  Individuals exhibiting superior fitness are more likely to be selected and subjected to diverse genetic operations, including mutation, crossover, and selection, resulting in the creation of a new generation of individuals or quantum circuits. This iterative process continues until the convergence criteria are met.  To illustrate the comprehensive encoding scheme encompassing various types of gates, we employ three qubits and allocate 10 bits per gene, as depicted in Figures \ref{encode} and \ref{encode-1}. 

\tikzstyle{s} = [rectangle, rounded corners, 
minimum width=3cm, 
minimum height=0.5cm,
text centered, 
draw=black, 
fill=red!30]

\tikzstyle{description} = [rectangle, 
minimum width=2cm, 
minimum height=1cm, 
text centered, 
text width=2cm, 
draw=black, 
fill=orange!30]

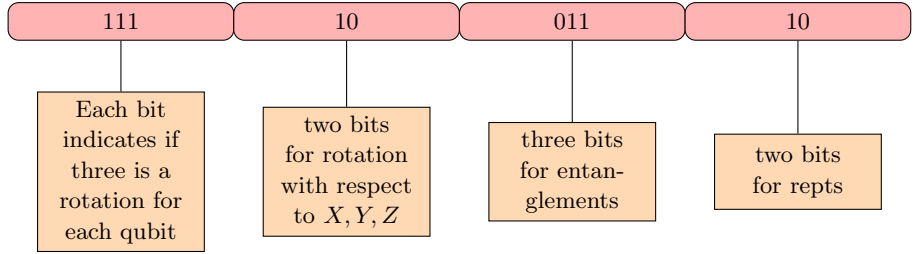
\begin{figure}[htbp]
\centering
\begin{tikzpicture}[node distance=2cm]
\node(2)[s,right of=1, xshift=1cm]{111};
\node(3)[s,right of=2, xshift=1cm]{10};
\node(4)[s,right of=3, xshift=1cm]{011};
\node(5)[s,right of=4, xshift=1cm]{10};

\node(2e)[description,below of=2]{Each bit indicates if three is a rotation for each qubit};
\node(3e)[description,below of=3]{two bits for rotation with respect to $X, Y, Z$};
\node(4e)[description,below of=4]{three bits for entanglements};
\node(5e)[description,below of=5]{two bits for repts};

\draw [-] (2) -- (2e);
\draw [-] (3) -- (3e);
\draw [-] (4) -- (4e);
\draw [-] (5) -- (5e);
\end{tikzpicture}
\caption{Example for the encoding scheme:  the number of qubits is 3.    the first three bits for 3 rotation gates, and next two bits indicates the rotation with respect to $X, Y, Z$, followed 3 bits for possible entanglements,  the last two bits are for repetitions. }
\label{encode-1}
\end{figure}

\subsection{Fitness functions}\label{sec3-2}

Fitness, in this context, serves as the objective function or cost function. It takes a solution as input and yields the fitness score for that solution. Based on the fitness functions proposed in \cite{ARG21}, we introduce three fitness functions below (\ref{FitnessFunc}), our aim is twofold: to {\bf maximize accuracy} and {\bf minimize both of local and non-loccal gate costs}. In tackling multi-objective problems, a common strategy involves identifying the Pareto front \cite{ADB14, Miettinen1999, Deb2002}, which includes selecting high-dominance points using crowd distance techniques. Recognizing that different gates may have distinct costs \cite{ARG21, CC2022, LLK06}, we distinguish CNOT gates from local gates and introduce three fitness optimization functions. The resulting three-dimensional Pareto fronts offer valuable insights for devising cost-effective trade-off strategies. Accuracy, in this context, denotes the mean accuracy achieved on the provided test data and labels, using the SVM model generated by a quantum circuit. Gate costs are computed by counting the sequence of basic physical operations required for quantum computer implementation.

\begin{equation}\label{FitnessFunc} 
\left\{
\begin{array}{lr}
\text{(maximize)} \text{Fitness 1} = \text{prediction accuracy}\\ 
\text{(minimize)} \text{Fitness 2 } =\text{ local gates}=\text{Rgate} + \text{Hgate} \\
 \text{(minimize)} \text{Fitness 3 } = \text{CNOT gate}
\end{array}
\right. 
\end{equation}

We choose not to introduce weights to the objective functions;  such weighting could exert influence on both accuracy and gate count, potentially affects the optimization process. Our fitness function is designed for optimizing two key aspects: the maximization of classification accuracy and the minimization of the variational circuit's complexity for both local and non-local gates. In the pursuit of this optimization, we undertake a rigorous evaluation process. This process involves K-fold cross validation. Utilizing the quantum circuit along with the training set, we compute the classifier, employing the quantum kernel SVM. Subsequently, we measure the model's performance by averaging its accuracy over the test set. 
%------------------ Section 4-----------------------

\section{Experiments with two datasets} \label{sec5}

\subsection{Experimental Procedure}
\begin{figure}
\centering

\tikzstyle{processA} = [rectangle, 
minimum width=1cm, 
minimum height=4cm, 
text centered, 
text width=2.1cm, 
draw=black, 
fill=yellow!30]

\begin{tikzpicture}[node distance=1cm]

\node (2) [processA, right of=1, xshift=1.7cm] {Data Preprocess (K-fold, population encoding,..)};
\node (3) [processA, right of=2,xshift=1.7cm] {NSGA-II Process (Crossover, Mutation)}; 
\node (4) [processA, right of=3,xshift=1.7cm] {Construct Quantum Feature Map and Compute Kernel Matrix};
\node (6) [processA, right of=4, xshift=1.7cm] {Compute Accuracy and others (Classical Support Vector Machine with Quantum Kernel)};
\node (8) [processA, right of=6, xshift=1.7cm] {NSGA-II Process (Sorting,..)};
\node (9) [processA, right of=8, xshift= 1.7cm] {Get Pareto Fronts};

\node (10) [decision, above of=4, xshift= 1.5cm, yshift= 3cm] {K-fold Repetition};
\node (11) [decision, below of=4, xshift= 1.5cm, yshift= -3cm] {NSGA-II Repetition};

\draw [->] (2) -- (3);
\draw [->] (3) -- (4);
\draw [arrow] (4) -- (6);
\draw [arrow] (6) -- (8);
\draw [arrow] (8) -- (9);

\draw [arrow] (8) |- (11);
\draw [arrow] (11) -| (3);
\draw [arrow] (9) |- (10);
\draw [arrow] (10) -| (2);

%\draw [arrow] (8) -- node[anchor=east] {yes} (9);
%\draw [arrow] (8) -- node[anchor=south] {no} (10);
%\draw [arrow] (10) |- (2);
%
%\draw [arrow] (4) -- (11);
%\draw [arrow] (11) -- (12);
%\draw [arrow] (12) -- (5);

%\draw [arrow] (dec1) -- node[anchor=east] {yes} (pro2a);
%\draw [arrow] (dec1) -- node[anchor=south] {no} (pro2b);
%\draw [arrow] (pro2b) |- (pro1);
%\draw [-] (pro2a) -- (out1);
%\draw [arrow] (out1) -- (stop);
\end{tikzpicture}
  \caption{Experimental procedure with NSGA-II and quantum feature map}
  \label{procedure1}
\end{figure}
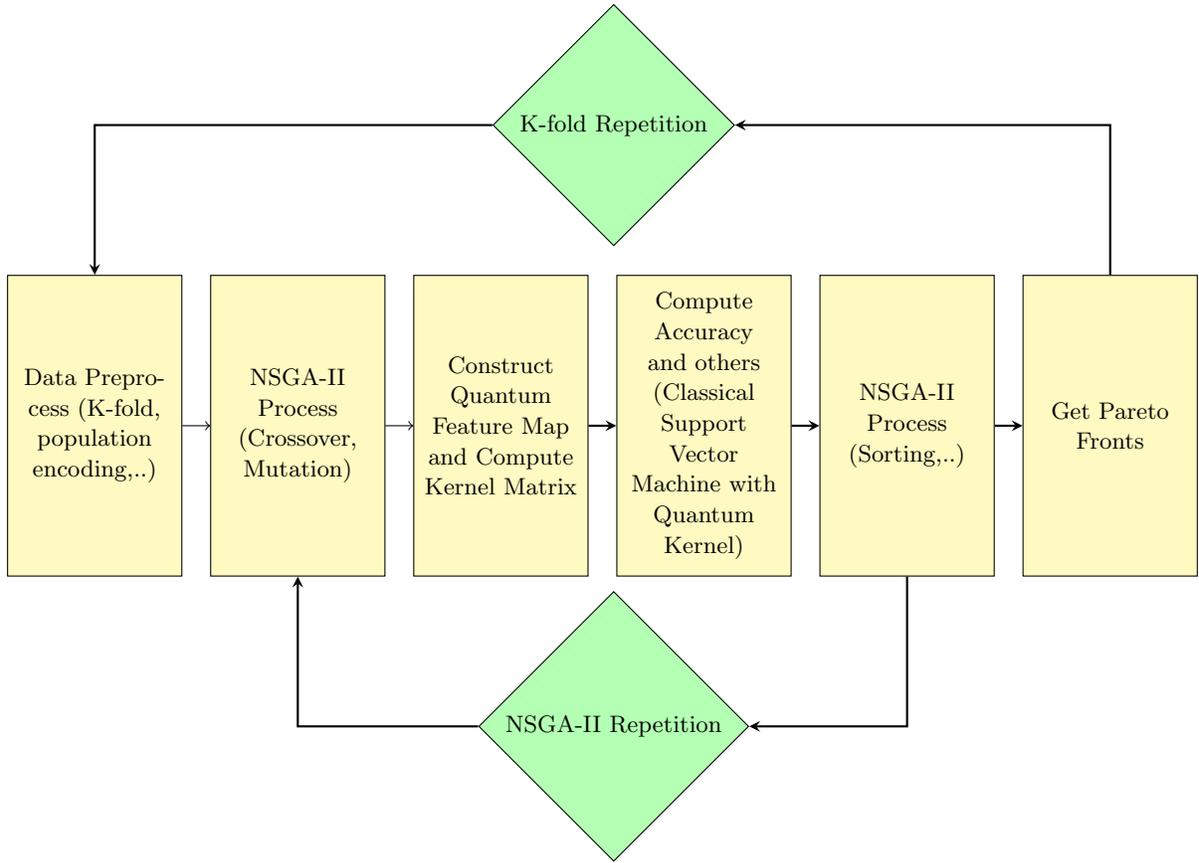

The experiment leverages a combination of high-dimensional NSGA-II (Non-dominated Sorting Genetic Algorithm II) and quantum feature mapping for achieving higher accuracy. We use two datasets: one is the breast cancer dataset containing 30 features, and the other is the Iris dataset with 4 features.  To intensify the challenge for achieving high accuracy, we opt not to employ principal component analysis (PCA) for dimension reduction.  Once the each individual gene is encoded for quantum feature maps as in Section \ref{sec4}, we proceed to genetic algorithm operators such as selection, crossover, and mutation to generate the next generation of individuals. Then we compute the kernel of a quantum SVM while training a classifier with the provided training dataset. In each loop, we evaluate the accuracy of the classifier using the test dataset and assess the effective size of the circuit in terms of the number of local and entangling gates it contains.  The procedure is organized into a series of interconnected steps represented by distinct rectangular blocks as in Figure \ref{procedure1}. Special cares have been taken to avoid over fitting by implementing K-fold cross validation and comparison with baseline model's performance.   

\begin{itemize}
    \item \textbf{Data Preprocess}: The initial step involves the preprocessing of the data using a K-fold strategy, ensuring that the dataset is divided into subsets for training and testing. This is a common technique in machine learning to assess the model's performance and robustness. In addition, individual gene in populations are encoded for quantum feature maps. 

    \item \textbf{NSGA-II Process (Crossover, Mutation)}: NSGA-II, a multi-objective optimization algorithm, is applied to the problem. It encompasses various genetic operators such as crossover and mutation, designed to evolve a population of potential solutions to improve their quality with respect to multiple objectives.

    \item \textbf{Construct Quantum Feature Map and Compute Kernel Matrix}: Here, a quantum feature map is constructed, which is a crucial element of quantum computing used to transform classical data into a quantum state. The procedure also involves computing a kernel matrix, which is a fundamental concept in machine learning for capturing data relationships.

    \item \textbf{Classical Support Vector Machine to Compute Accuracy}: A classical Support Vector Machine (SVM) with the quantum kernel is employed to compute the accuracy of the model's predictions, which improves the effectiveness of the powerful classical method in machine learning.  

    \item \textbf{NSGA-II Process (Sorting)}: Another iteration of the NSGA-II process is performed, potentially refining the solutions generated earlier.

    \item \textbf{Get the Pareto Fronts}: The final step aims to obtain the Pareto fronts, which are solutions that represent the optimal trade-offs between conflicting objectives. These Pareto fronts are often used to guide decision-making in multi-objective optimization problems.
\end{itemize}

\subsection{Breast cancer dataset}

We initiate our analysis with the breast cancer dataset, sourced from Sklearn \cite{PVG11, Zwitter1998}. This dataset contains measurements of breast tissue derived from a medical imaging technique, including various measurements related to cell nuclei. The primary objective is to ascertain whether a tumor is benign (harmless) or malignant (cancerous and dangerous). The breast cancer dataset, imported from scikit-learn, comprises 569 samples, encompassing 30 real, positive features. These features encompass attributes related to cancer mass, such as mean radius, mean texture, mean perimeter, and more. Among these samples, 212 are labeled as "malignant," while 357 are labeled as "benign."

Subsequently, we randomly select a number of features equal to the number of quantum circuits for training.  To ensure robustness in our analysis, we randomly selected 50 feature combinations for each qubit size. Having our training and testing datasets prepared, we proceed with the previously outlined procedure to compute the training and testing kernel matrices.  We start with initial random population of 100 individuals represented as binary chains, each comprising $N+\binom{N}{2} +4$ bits, where $N$ represents the number of qubits. We use 20\% crossover and mutation rate. To get a more realistic accuracy, we use the 5-fold cross-validation methods to have 5 different training and test subsets and then average accuracy and other outputs.  We use a combination of generation number and accuracy as stopping criterion.  These quantum kernels can subsequently be integrated into classical kernel methods.  The NSGA-II typically run only 10 generations to achieve higher accuracy than the classical SVMs.

%Figure \ref{fig:boundary} illustrates the data points and prediction boundaries resulting from the quantum kernel SVM in a two-qubit scenario, while Figure \ref{fig:confusion} presents the corresponding confusion matrix. The confusion matrix is a pivotal tool for evaluating the performance of a classification algorithm. Together, these elements furnish an estimate of the quantum kernel matrix, which can be harnessed as a kernel for support vector classification.
%
%\begin{figure}
%    \centering
%    \begin{subfigure}[t]{0.30\textwidth}
%        \centering
%        \includegraphics[width=\linewidth]{DecisionBoundary22-6.png} 
%        \caption{Prediction boundary} \label{fig:boundary}
%    \end{subfigure}
% %   \hfill
%    \begin{subfigure}[t]{0.30\textwidth}
%        \centering
%        \includegraphics[width=\linewidth]{confusion22-6.png} 
%        \caption{Confusion matrix} \label{fig:confusion}
%    \end{subfigure}
%    \caption{Prediction boundary and confusion matrix}
%\end{figure}
%
%
We conducted a comprehensive comparison of various kernels and quantum feature mapping techniques. Specifically, we assessed the performance of four well-established classical kernels (Linear, Poly, RBF, Sigmoid) and Logistic Regression, Naive Bayes K-Neighbors and Decision Tree.  The results, summarized in Table \ref{table_compare1}, showcase the average best accuracy achieved by each prediction method across different feature and qubit configurations.

Our findings reveal that, in comparison to classical kernel methods, the quantum kernel generated by the genetic algorithm consistently outperforms all classical kernels with the dataset.  These results underscore the suitability of the quantum kernel for the given dataset, highlighting its potential as a powerful tool for enhancing classification tasks.

\begin{ruledtabular}
\begin{table}[htbp]
    \centering
    \caption{Accuracy comparison for prediction methods with breast cancer dataset. The accuracy is calculated as the average of  the best accuracies for 50 randomly selected features.} 
    \begin{tabular}{lccccccccc}
      \multirow{2}{*}{Prediction Method} & \multicolumn{9}{c}{Number of features/qubits }   \\
                             &  2        & 3  & 4     &  5     & 6    &  7  &  8  &  9 &  10 \\
      \\[-1.2em]
      \hline
      \\[-1.2em]
SVM (Linear) & 0.85 & 0.89 & 0.91 & 0.92 & 0.93 & 0.94 & 0.95 & 0.95 & 0.95 \\

SVM (Poly) & 0.85 & 0.9 & 0.92 & 0.93 & 0.93 & 0.94 & 0.94 & 0.95 & 0.95 \\

SVM (RBF) & 0.85 & 0.89 & 0.91 & 0.92 & 0.93 & 0.94 & 0.94 & 0.95 & 0.95 \\

SVM (Sigmoid) & 0.8 & 0.83 & 0.84 & 0.85 & 0.87 & 0.88 & 0.88 & 0.88 & 0.9 \\

Logistic Regression & 0.85 & 0.89 & 0.91 & 0.92 & 0.93 & 0.94 & 0.94 & 0.94 & 0.95 \\

Naive Bayes & 0.84 & 0.89 & 0.9 & 0.91 & 0.91 & 0.92 & 0.92 & 0.92 & 0.93 \\

K-Neighbors & 0.84 & 0.88 & 0.9 & 0.92 & 0.93 & 0.93 & 0.94 & 0.94 & 0.95 \\

Decision Tree & 0.85 & 0.88 & 0.9 & 0.91 & 0.91 & 0.92 & 0.92 & 0.92 & 0.93 \\
      \\[-1.2em]
      \hline
      \\[-1.2em]
      SVM (Quantum kernel)          & 0.87  & 0.92 & 0.94  & 0.95 & 0.96 & 0.96 & 0.97& 0.97& 0.97\\
    \end{tabular}
   \label{table_compare1}
\end{table}
\end{ruledtabular}

\begin{ruledtabular}
\begin{table}[htbp]
    \centering
    \caption{Gate numbers of Quantum feature map with breast cancer data. The gate numbers are calculated as the average gates with the best accuracies of 50 randomly selected features.}
    \begin{tabular}{lccccccccc}
      \multirow{2}{*}{Gate} & \multicolumn{9}{c}{Number of features/qubits }   \\
                              &  2 & 3        & 4  & 5     &  6          & 7    & 8   & 9  & 10 \\
      \\[-1.2em]
      \hline
      \\[-1.2em]
     Local gates                &  9.11  &   12.1          &   15.88   &  18.07 &       22.23       &      29.29        & 33.1  &38.63 & 41.79\\
    %   \\[-1em]
     CNOT gates                & 2.07     & 4.67              & 8.5    & 11.71  &         17.66     &         26.77  &  33& 41.03& 46.66\\
       \end{tabular}
    \label{table_compare2}
\end{table}
\end{ruledtabular}

Furthermore, Table \ref{table_compare2} provides valuable insights into the average quantities of local and CNOT gates associated with our quantum circuits. It's noteworthy that in our experiments with the dataset, we observe a reasonable proportionality between the number of CNOT gates and the number of local gates. This finding deviates from the observations made in \cite{ARG21, CC2022}, where entanglement gates were predominantly suppressed.

The discrepancy may be attributed to the optimization functions employed in \cite{ARG21}, which introduce weighting factors to strike a balance between the relative significance of the two performance metrics. As detailed in \cite{ARG21}, assigning a high weight to accuracy can lead to a convergence into a single individual and lose the necessary genetic diversity crucial for minimizing the quantum circuit's size throughout evolution.

\subsection{Iris dataset}\label{sec5-2}

\begin{ruledtabular}
\begin{table}[htbp]
    \centering
    \caption{Accuracy comparison for the kernel methods with Iris dataset. The best accuracy for each prediction method with the different features}
    \label{table_compare11}
\begin{tabular}{lccc}
    \multirow{2}{*}{Prediction methods} & \multicolumn{3}{c}{Number of features/qubits} \\
    & 2 & 3 & 4 \\
    \hline
   SVM (Linear) & 0.93 & 0.96 & 0.97 \\
   SVM (Poly) & 0.93 & 0.96 & 0.98 \\
   SVM (RBF) & 0.94 & 0.95 & 0.95 \\
   SVM (Sigmoid) & 0.90 & 0.93 & 0.95 \\
  Logistic Regression  & 0.91 & 0.94 & 0.95 \\
 Naive Bayes  & 0.91 & 0.94 & 0.95 \\
 K-Neighbors  & 0.91 & 0.94 & 0.95 \\
 Decision Tree  & 0.92 & 0.96 & 0.97 \\
 \\[-1.2em]
      \hline
      \\[-1.2em]   
 SVM(Quantum Kernel) & 0.94 & 0.98 & 0.99 \\
\end{tabular}

\end{table}
\end{ruledtabular}

\begin{ruledtabular}
\begin{table}[htbp]
    \centering
    \caption{Gate numbers for the quantum kernel method with Iris data. Gate numbers with the best accuracy}
    \label{table_compare22}
     \begin{tabular}{lccc}
        \multirow{2}{*}{Gate} & \multicolumn{3}{c}{Number of features/qubits} \\
        & 2 & 3 & 4 \\
        \\[-1.2em]
        \hline
        \\[-1.2em]
        Local gates  & 4.07 & 5.05 & 8.4\\
        CNOT gates  & 0.47 & 0.6 & 3.6\\
        \\[-1.2em]
        \end{tabular}
\end{table}
\end{ruledtabular}

%
%\begin{figure}[htbp]
%    \centering
%    \begin{subfigure}[t]{0.30\textwidth}
%        \centering
%        \includegraphics[width=\linewidth]{Pateto011.png} 
%        \caption{Pareto optimal front} \label{fig:timing11}
%    \end{subfigure}
% %   \hfill
%    \begin{subfigure}[t]{0.30\textwidth}
%        \centering
%        \includegraphics[width=\linewidth]{Pateto012.png} 
%        \caption{Pareto optimal front (accuracy with local gates)} \label{fig:timing22}
%    \end{subfigure}
%%    \vspace{1cm}
%    \begin{subfigure}[t]{0.30\textwidth}
%    \centering
%        \includegraphics[width=\linewidth]{Pateto013.png} 
%        \caption{Pareto optimal front (accuracy with CNOT gates)} \label{fig:timing33}
%    \end{subfigure}
%    \caption{Three and two dimensional Pareto optimal fronts for Iris data}
%\end{figure}
Next, our attention turns to the Iris dataset, which has three distinct types of irises: Setosa, Versicolour, and Virginica. This dataset includes measurements of petal and sepal length, organized in a 150x4 array format, where the rows represent individual samples, and the columns encompass Sepal Length, Sepal Width, Petal Length, and Petal Width.

Given the dataset's relatively limited feature set of only 4 features, instead of random feature selection, we opt to evaluate the performance with 2, 3, and 4 features individually. In fact, this equates to a total of 6 features when using 2 qubits, 4 features with 3 qubits, and a single feature represented by 4 qubits. Subsequently, we execute a consistent procedure, measuring accuracy and facilitating a comparative analysis of quantum SVM kernels with classical SVM kernels and and Logistic Regression, Naive Bayes, K-Neighbors and Decision Tree.

We employ the same procedure as applied to the breast cancer dataset to perform a comparative analysis of various kernels, including those generated from quantum feature maps. 
Having our training and testing datasets prepared, we proceed with the previously outlined procedure to compute the training and testing kernel matrices.  We use 20\% crossover and mutation rate. To get a more realistic accuracy, we use the 5-fold cross-validation methods to have 5 different training and test subsets and then average accuracy and other outputs. The NSGA-II typically run only 20 generations to achieve higher accuracy than the classical SVMs.   Table \ref{table_compare11} showcases the highest accuracy achieved by each prediction method across varying numbers of features or qubits.

When assessing the classical kernel methods, it becomes evident that the quantum kernel generated by a genetic algorithm either matches or surpasses the performance of classical kernels with this dataset. Table \ref{table_compare22} provides insights into the corresponding counts of local and CNOT gates utilized within the quantum circuits. Notably, for the Iris dataset, it is observed that the numbers of CNOT gates are small compared to the breast cancer dataset. This phenomenon may be attributed to the effectiveness of local gates in achieving high accuracy, making the inclusion of entanglements unnecessary in this scenario.

%Figure  \ref{fig:timing11}, \ref{fig:timing22}, \ref{fig:timing33} depict the Pareto-optimal fronts in three and two dimensions for one instance.  The Pareto fronts  explicitly show the strategy to achieve a high accuracy while keeping the gate costs low. 

\subsection{Data separability and non-local gates}\label{sec5-3}

In this section, our objective is to look into the impact of data complexity, specifically, data separability, on the prevalence of entanglement circuits. The efficacy of a SVM classifier is contingent on two critical factors: the classifier model itself and the degree of separability or complexity inherent in the datasets under consideration. This characterization of data complexity often takes into account various factors, such as the data distribution, as well as estimates of the shape and size of the decision boundary \cite{lorena2019}.

For the sake of comparison and a quantitative assessment of dataset separability, we have chosen three distinct indices as measures. These indices provide valuable insights into the separability of the datasets and are instrumental in our analysis.

The first index is the separability index (SI) measure which estimates the average number of instances in a dataset that have a nearest neighbor with the same label. Since this is a fraction, the index varies between 0-1 or 0-100\%.  Another separability measure, based on the class distance or margin is the hypothesis margin index (HMI), introduced in \cite{GNT2004}. It measures the distance between an object’s nearest neighbor of the same class (near-hit) and a nearest neighbor of the opposing class (near-miss) and sums over these.

$$
\begin{array}{r}
\theta_{P}(x)=\frac{1}{2}(\|x-\operatorname{nearmiss}(x)\|- \|x-\operatorname{nearhit}(x)\|)
\end{array}
$$
where nearhit $(x)$ and nearmiss $(x)$ denote the nearest point to $x$ in $P$ with the same and different label, respectively.  Therefore,  large hypothesis-margin ensures large sample-margin, and the dataset has high separability.  Margins occupy a pivotal position in contemporary machine learning research, serving as essential metrics to gauge the classifier's confidence during decision-making. These margins serve a dual purpose: they are employed to establish theoretical generalization bounds and provide valuable insights for algorithm design. It is imperative to note that as the near-miss distance increases and near-hit values decrease, the hypothesis margin index proportionally expands.  

The third index is the distance-based separability index (DSI) which was recently introduced in \cite{Guan22}. They identified scenarios where different data classes are blended within the same distribution as particularly challenging for classifiers to disentangle. Prior to introducing the DSI (Data Separability Index), we first introduce two fundamental components: the intra-class distance (ICD) and the between-class distance (BCD) sets. These components are instrumental in the computation of the DSI.  Suppose $X$ and $Y$ have $N_x$  and $N_y$ data points, respectively. The intra-class distance (ICD) set $\{d_x\}$ is a set of distances between any two points in the same class $(X)$, as: $\{d_x\}=\{ \|x_i-x_j\|_2 | x_i,x_j\in X;x_i\neq x_j\}$.  The between-class distance (BCD) set $\{ d_{x,y}\}$ is the set of distances between any two points from different classes $(X \, and \, Y)$, as $\{ d_{x,y} \}=\{ \|x_i-~y_j\|_2 \, |\, x_i\in X;y_j\in Y \}$. The metric for all distances is \textit{Euclidean} $(l^2\,\text{norm})$. We firstly introduce the computation of the DSI for a dataset containing only two classes $X$ and $Y$.  First, the ICD sets of $X$ and $Y$: $\{d_x\},\{d_y\}$ and the BCD set: $\{d_{x,y}\}$ are computed by their definitions.  The similarities between the ICD and BCD sets are then computed using the the Kolmogorov–Smirnov (KS) distance   $$
    s_x=KS(\{d_x\},\{d_{x,y}\}),\ \text{and}\ s_y=KS(\{d_y\},\{d_{x,y}\}).
    $$ 
Finally, the DSI is the average of the two KS distances:
    $$
    DSI(\{X,Y\})=\frac{(s_x+s_y )}{2}.
    $$   
 
\begin{ruledtabular}
\begin{table}[htbp]
    \centering
    \caption{Separability indexes of breast cancer data}
\begin{tabular}{lccccccccc}
    \multirow{2}{*}{Separability Index} & \multicolumn{9}{c}{Number of features/qubits } \\
    & 2 & 3 & 4 & 5 & 6 & 7 & 8 & 9 & 10 \\
        \hline
    \\[-1.2em]
    SI  & 0.78 & 0.82& 0.84   & 0.87    & 0.88   & 0.88& 0.88 & 0.89 & 0.90\\
    HMI & 4.18 & 5.82 & 7.47 & 8.53   & 10.82 & 10.75 & 11.61 & 12.46 & 13.36 \\
    DSI & 0.27 & 0.28 & 0.32 & 0.34   & 0.38   & 0.35 & 0.35 & 0.36 & 0.37 \\
\end{tabular}
    \label{table_compare33}
\end{table}
\end{ruledtabular}
 
\begin{ruledtabular}
\begin{table}[htbp]
    \centering
    \caption{Separability indexes of Iris data}
    \begin{tabular}{lccc}
        \multirow{2}{*}{Separability Index} & \multicolumn{3}{c}{Number of features/qubits} \\
        & 2 & 3 & 4 \\
        \hline
        SI  & 0.9 & 0.94 & 0.95 \\
        HMI & 8.29 & 10.46 & 12.19 \\
        DSI & 0.78 & 0.80 & 0.82 \\
    \end{tabular}
\label{table_compare44}
\end{table}
\end{ruledtabular}

For the breast cancer dataset, for each qubit size, we randomly select 50 combinations of the features with this size,  Tables \ref{table_compare33} presents the corresponding average numbers of the three separability indexes along with each number of the features/qubits.  For comparison, we calculated the corresponding average three separability indexes for three qubits/features  in Table \ref{table_compare44} for the Iris dataset (in fact, there are only 6 features with 2 qubits, and 4 features with 3 qubits and one feature with 4 qubits). This comparison of the separability indexes for the two datasets may explain why we need more CNOT gates for the breast cancer dataset. From the definitions, it is clear that \textit{the higher the three indexes are, the easier it is to separate a dataset.}   Clearly the separability indexes for the Iris dataset are larger than those of the breast cancer dataset, which explains why the quantum feature maps for the breast cancer dataset needs more CNOT gates to achieve high accuracy.   For the Iris data, which has much higher separability indexes, and it seems that the CNOT gates are almost zero. In addition, for the breast cancer dataset, we see that the higher dimensions of features/qubits,  the more CNOT gates are needed.  This indicates that entanglements may be only needed for lower separability indexes and high dimensional qubits.   

%---------------- Section 5------------------------
\section{Conclusion and discussion} \label{sec6}
Entanglement, a fundamental concept in quantum mechanics, serves as a critical factor in empowering quantum computing. In this paper, we explore the optimization of entanglement gates within quantum feature maps for quantum support vector machines using a genetic algorithm (specifically NSGA-II). Our approach involves using a multi-objective genetic algorithm with three fitness functions to achieve two key objectives simultaneously: maximizing classification accuracy and minimizing the gate costs associated with the quantum feature map's circuit. These gate costs in the optimization are separated into two distinct functions: one for local gates and another specific to non-local gates. 

As a result, we acquire a collection of high dimensional non-dominated points that unveil a quantum circuit configuration for the quantum kernel method. The analysis of this three-dimensional set of non-dominated points yields insights into the influence of entanglement gates on the performance of quantum support vector machines. The experimental results in this work shed light on the composition of quantum kernel maps for support vector machines, indicating a proportional inclusion of CNOT gates to promote entanglement. These findings complement prior findings in \cite{ARG21,CC2022}, where CNOT gates were significantly suppressed. Our experiments also suggest that higher-dimensional qubits may need a greater number of CNOT gates to achieve high prediction accuracy.

This study also employs three separability indexes to investigate the impact of data on the configuration of feature maps in quantum support vector machines. Our experiments, while still needs theoretical analysis,  reveal that, for the same dataset, higher-dimensional qubits may require an increased number of CNOT gates to achieve improved prediction accuracy. Moreover, datasets with less separability indexes tend to benefit from additional CNOT gates to attain high prediction accuracy. These findings highlight the utility of various separability indexes in aiding the design of efficient quantum circuits for achieving higher accuracy.

It's noteworthy that the current documentation for quantum kernels on https://qiskit.org/ lacks guidance on parameter selection. For instance, the entanglement parameter in ZZFeatureMap offers three options: linear, full, and circular. The insights from this paper can serve as a valuable resource for making informed decisions regarding these parameters based on data analysis.
 
The experiments of data separability reveal an interesting trend: less data separability necessitates the utilization of more CNOT gates to achieve accurate predictions. This observation can be useful for the design of more efficient quantum circuits for machine learning applications. Future research could focus on refining indexes and metrics for assessing data separability, potentially unveiling the presence of a quantum advantage more effectively.

The potential advantages of using quantum kernel estimation to enhance machine learning applications are still under exploration \cite{Schuld2022}. There is no universally superior model in machine learning; the effectiveness of a model often depends on the distribution and nature of the training data. In addition to the separability indexes we discussed in this paper, various metrics have been introduced to measure the performance and efficiency of quantum kernels within machine learning applications.  One such methodology, developed by \cite{huang2021}, centers on the geometric difference of data within feature spaces. This metric provides valuable insights into the variations in prediction errors \cite{Di2023}. 

The Pareto fronts generated from multiple fitness functions for genetic algorithms illustrate the methodology for attaining high accuracy while concurrently minimizing gate costs and other associated properties. While we only examine the gate costs associated the maximum accuracy in this paper,  it would be interesting to carefully study cost-effective trade-off strategies of different circuit configurations in a future work, particularly in scenarios where multiple objectives may be in conflict with each other. Further exploration of high-dimensional Pareto-optimal fronts and their projections will provide additional insights into the substantial influence of entanglements on quantum kernel accuracy. 

%----------------------------------------------

\section{Acknowledgments}

The author would like to thank Allison Brayro, Nao Yamamoto, Sergio Altares-López, Juan José García-Ripoll, Angela Ribeiro, Nam Nguyen for reading an earlier draft of this paper and providing comments. 

%---------------- References---------------------

\section{Declarations}

Funding:   No funding is associated with the paper.
 
Availability of data and materials:  The datasets used in paper are publicly available in the Python package sklearn.

\end{document}